\documentclass[twocolumn,pre]{revtex4-2}
\usepackage{graphicx}
\usepackage{hyperref}
\usepackage[utf8]{inputenc}
\usepackage{float}
\usepackage{bbm}
\usepackage[usenames,dvipsnames]{xcolor}
\usepackage{amsmath,amsfonts,amssymb}

\newcommand{\mitch}[1]{\color{ForestGreen}{#1}\color{black}}

\begin{document}
\title{Optimal network geometry detection for weak geometry}
\author{Riccardo Michielan, Clara Stegehuis}
\date{\today}

\begin{abstract}
    Network geometry, characterized by nodes with associated latent variables, is a fundamental feature of real-world networks. Still, when only the network edges are given, it may be difficult to assess whether the network contains an underlying source of geometry. 
    This paper investigates the limits of geometry detection in networks in a wide class of models that contain geometry and power-law degrees, which include the popular hyperbolic random graph model. We specifically focus on the regime in which the geometric signal is weak, characterized by the inverse temperature $1/T<1$. 
    We show that the dependencies between edges can be tackled through Mixed-Integer Linear Problems, which lift the non-linear nature of network analysis into an exponential space in which simple linear optimization techniques can be employed. This approach allows us to investigate which subgraph and degree-based statistic is most effective at detecting the presence of an underlying geometric space. Interestingly, we show that even when the geometric effect is extremely weak, our Mixed-Integer programming identifies a network statistic that efficiently distinguishes geometric and non-geometric networks.
\end{abstract}

\maketitle

\section{Introduction}
Network geometry, characterized by nodes with associated latent variables, is a crucial feature of real-world networks~\cite{boguna2021,serrano2003}. It accounts for various network properties such as scale-invariance, high clustering, and overlapping community structures. Random graph models that integrate geometric properties typically assign each node a position in an underlying geometric space, ensuring that proximate nodes are more likely to connect. This framework encompasses the concept of similarity or homophily, as seen in social networks, where similar individuals are more likely to connect.

Numerous geometric random graph models have been proposed to represent real-world networks~\cite{krioukov2010,bringmann2016,deijfen2013scale,serrano2008,kaminski2020}. The simplest geometric models are based on an underlying Euclidean space~\cite{penrose2003,dall2002} and connect all nodes within a specified radius. However, these models do not capture the heavy-tailed degree distributions which are often observed in real-world networks. To address this, one can either transition the underlying geometric space from Euclidean to hyperbolic space~\cite{krioukov2010}, or enhance Euclidean models by incorporating an additional hidden variable for each node that encodes the expected degree of each node~\cite{serrano2008,bringmann2016,kaminski2020}.

Recent evidence suggests that the geometric influence on network connections is `weak' in many networks~\cite{kolk2024}: While nearby vertices are more likely to connect than distant vertices, in the weak geometric regime, connections between distant vertices are relatively likely. This scenario, also referred to as the `high-temperature regime', where the inverse temperature $1/T<1$, has received limited attention in recent years, with notable exceptions being~\cite{kolk2024,kolk2024_random}. 

Although network geometry is a natural assumption, network samples often only include the edges, without the positions in the underlying geometric space. Identifying whether a given network has an underlying latent hyperbolic geometric space has therefore been recognized as a significant open problem~\cite{boguna2021,smith2019}. Specifically, when the degree distribution is heavy-tailed, simple statistics such as triangle counts are ineffective even in the strong geometric regime~\cite{michielan2022}. This raises the question of whether evaluating the network geometry using other edge-based observables is possible. In the strong geometry regime, a specific form of triangles can distinguish between geometric and non-geometric models~\cite{michielan2022}. However, in the weak geometry regime, much less is known. Basic statistics such as clustering coefficients fail to differentiate between geometric and non-geometric networks when the inverse temperature $1/T < 0.5$~\cite{kolk2024}. 

This paper investigates the limits of geometry detection. It investigates when often-used plain subgraph counts can distinguish geometric and non-geometric networks effectively, and when more intricate statistics are necessary.  Furthermore, while previous approaches often introduce specific statistics and show that they are effective~\cite{michielan2022,kolk2024,bubeck2016,funke2006}, here we introduce a general approach that can effectively optimize and seek the best possible statistic that is possible to distinguish geometric and non-geometric networks with the highest expected difference.

The contributions of this paper are threefold. First of all, we show that weak geometry is \emph{always} detectable, that is, for all temperatures $1/T>0$. Secondly, we present a flexible Mixed-Integer Linear Programming approach capable of characterizing the behavior of various local network statistics. This method searches for the most effective degree-based subgraph statistic to differentiate between geometric and non-geometric graphs. The fact that the approach is based on Mixed Integer Linear Programs makes it easy to use state-of-the-art solvers such as Gurobi~\cite{gurobi} to optimize statistics to detect geometry efficiently. While network connection probabilities are typically non-linear, we show that by lifting to an exponential space and introducing binary variables, network analysis can be transformed into a linear optimization problem involving binary variables. Finally, we show that subgraph count statistics alone can only detect weak geometry up to a certain threshold. However, by combining degree and subgraph information, it is possible to detect weak geometry at any level of $1/T > 0$.

\paragraph*{Organization of the paper.} We introduce the network model we analyze in Section~\ref{sec:model}. We then introduce the Mixed-Integer Linear Programming approach that we use for $1/T>0$ and $1<T<0$ in Sections~\ref{sec:motifhightemp} and~\ref{sec:motiflowtemp} respectively. Section~\ref{sec:optmotif} then investigates the optimal subgraph statistic to detect (weak) network geometry and the inherent temperature threshold below which no subgraph can detect weak geometry. Section~\ref{sec:optweightedmotif} then shows that adding degree information allows to detect weak geometry at any $1/T>0$. Section~\ref{sec:conclusion} finalizes the paper with a conclusion and discussion.

\section{Model}\label{sec:model}
We now introduce the geometric model that we consider, the geometric inhomogeneous random graph~\cite{bringmann2016}, which generalizes the hyperbolic random graph~\cite{krioukov2010,komjathy2020} and the $\mathbb{S}^1$ model~\cite{serrano2008}.
In this model, every vertex $i$ is equipped with a weight $w_i$, and a $d$-dimensional position $x_i$. The weights $(w_i)_{i\in[n]}$ are sampled independently from the power-law distribution
\begin{equation}\label{eq:pl}
    \rho(x) = Cx^{-\tau},
\end{equation}
for some $\tau\in(2,3)$. The positions $(x_i)_{i\in [n]}$ are sampled independently and uniformly on the $d$-dimensional torus. Given the weights and positions, the connection probability is given by
\begin{equation}\label{eq:conproblow}
 p_{ij} = \min\Big(1 ,\left(\frac{w_i w_j}{n \mu |x_i - x_j|^d}\right)^{1/T}\Big),
 \end{equation}
 in the low-temperature regime $1/T>1$. Here $\mu$ denotes the average weight of the distribution~\eqref{eq:pl}.

In the high-temperature regime $1/T > 1$ the connection probability is given by
\begin{equation}\label{eq:conprobhigh} p_{ij} = \min\Big(1 , \frac{w_i w_j}{n \mu |x_i - x_j|^{d/T}}\Big)
\end{equation}

A non-geometric equivalent of this model is the Chung-Lu model~\cite{chung2002connected}, in which vertices are only equipped with weights, again sampled from~\eqref{eq:pl}. Then, the connection probability of nodes $i$ and $j$ is given by
\begin{equation}\label{eq:conprobcl}
    p_{ij} = \min\Big(1 , \frac{w_i w_j}{\mu n }\Big).
\end{equation}
This model is equivalent to letting $T\to\infty$ in~\eqref{eq:conprobhigh}.
For connection probabilities~\eqref{eq:conproblow},\eqref{eq:conprobhigh} and~\eqref{eq:conprobcl}, the expected degree of a vertex is equal to its weight. For high-degree vertices, the degrees concentrate quite well around the vertex weights, and $d_i=w_i(1+o_p(1))$~\cite{stegehuis2019,bringmann2016} (or $d_i=Kw_i(1+o_p(1))$ for some constant $K$~\cite{michielan2022}).

\section{Motif optimization: $1/T>1$ }\label{sec:motiflowtemp}
We first investigate the potential of plain subgraph counts to distinguish geometric and non-geometric models. To that end, we investigate the scaling of the subgraph count of a given subgraph $H =(V_H,\mathcal{E}_H)$ in the GIRG, $N^{GIRG}(H)$, similarly as in~\cite{michielan2024optimal}. 

We start with the low-temperature regime $1/T>1$. We show that all subgraphs typically appear on vertices with specific degrees and demonstrate how a mixed-integer linear program can determine these degrees. In particular, we show that a given subgraph $H$ on $k$ vertices is most likely to appear on vertices with weights $(n^{\alpha_i})_{i\in[k]}$ that have distances $(n^{\beta_{ij}})_{ij\in V_H\times V_H}$ for some $0\leq \alpha_i\leq1$ and $-1\leq \beta_{ij}\leq 0$. We will then show that for every subgraph $H$ one particular choice of $\mathbf{\alpha}$ and $\mathbf{\beta}$ dominates, indicating that subgraph $H$ is most likely to appear on vertices with those degrees and inter-distances. We therefore compute the expected number of copies of $H$ that appear on vertices with weights $(n^{\alpha_i})_{i\in[k]}$ that have distances $(n^{\beta_{ij}})_{ij\in V_H\times V_H}$.

We first compute the probability that $k$ randomly sampled vertices have weights proportional to $(n^{\alpha_i})_{i\in[k]}$.
   A vertex $v_i$ has weight proportional to $n^{\alpha_i}$ with probability proportional to $n^{(1-\tau)\alpha_i}$ due to~\eqref{eq:pl}. Therefore, the probability that $k$ vertices have weights proportional to $n^{\alpha_1},\dots,n^{\alpha_k}$ is proportional to
   \begin{equation}\label{eq:sampleweights}
       n^{(1-\tau)\sum_{i\in[k]}\alpha_i},
   \end{equation}
   as long as $\alpha_i\leq 1/(\tau-1)$ for all $i$. In fact, the highest degree among the $n$ independently sampled copies from~\eqref{eq:pl} scales with high probability as $n^{1/(\tau-1)}$, also called the natural cutoff~\cite{boguna2003class}.

   We then compute the probability that $k$ randomly sampled vertices have distances proportional to $(n^{\beta_{ij}})_{ij\in V_H\times V_H}$
Calculating this probability for all pairs of vertices $v_i,v_j$ is slightly more complicated than the computation for the weights, as the distances between vertices are dependent. For example, if $v_i$ and $v_j$ are close, and $v_k$ is close to $v_i$ as well, $v_k$ and $v_j$ cannot be too far from each other due to the triangle inequality. 
We therefore calculate the desired probability sequentially, starting from $v_1$. The probability that the distance between $v_2$ and $v_1$ is proportional to $n^{\beta_{\{1,2\}}}$ is proportional to $n^{d\beta_{\{1,2\}}}$. Indeed, the positions are sampled uniformly on the $d$-dimensional torus, and the region in which the distance to $v_1$ is proportional to $n^{\beta_{\{1,2\}}}$ is proportional to $n^{d\beta_{\{1,2\}}}$, and the volume of the $d$-dimensional torus is constant.
 
We then proceed to vertex $v_{3}$, and compute the likelihood of the distance from $v_3$ to $v_1$ and $v_2$ being proportional to $n^{\beta_{\{1,3\}}}$ and $n^{\beta_{\{2,3\}}}$ respectively. The area in which $v_3$ satisfies this is proportional to $\min(n^{d\beta_{\{1,3\}}},n^{d\beta_{\{2,3\}}})$. Iterating this procedure, combined with the fact that positions are sampled uniformly on the torus, yields that the probability that vertices $v_i$, $v_j$ have inter-distance $n^{\beta_{\{i,j\}}}$ for all $1\leq i<j\leq k$ is proportional to
\begin{equation}\label{eq:samplepositions}
    n^{d\sum_{j\in[k]}\min_{i:i<j}\beta_{\{i,j\}}}.
\end{equation}
Here we assume that $\beta_{ij}\in[-1/d,0]$, as the maximal distance on the torus is $1/2$, and the distance between the two closest points among $n\choose 2$ uniformly sampled pairs on the torus scales as $n^{-1/d}$ with high probability.

However, not all combinations of $\beta$ variables are allowed. As these variables denote distances, they should satisfy the triangle inequality. That is, for all $i,j,s$, the distance between $i$ and $j$ cannot be larger than the distance between $i$ and $s$ plus the distance between $j$ and $s$. As we only have the distances in orders of magnitude, this means that the order of magnitude of the distance between $i$ and $j$, $n^{\beta_{ij}}$ cannot be larger than the maximum of the order of magnitudes of the distance between $i$ and $s$, $n^{\beta_{is}}$,  and that between $j$ and $s$, $n^{\beta_{js}}$. In other words,~\eqref{eq:samplepositions} only holds as long as
\begin{equation}
    \beta_{\{i,j\}} \leq \max(\beta_{\{i,s\}},\beta_{\{s,j\}})
        \quad \forall i\neq j \neq s \in [k].
\end{equation}

While in general, the probability that subgraph $H$ forms on $k$ randomly selected vertices is difficult to compute, as the weights and distances results in correlations between the presences of edges, conditionally on the weights and distances, the connection probabilities are independent. Therefore, by~\eqref{eq:conproblow}, the probability that vertices with weights proportional to $n^{\alpha_i}$ $i\in[k]$ and distances proportional to $n^{\beta_{\{i,j\}}}$ $i,j\in[k]$ form subgraph $H$ is proportional to
\begin{align}\label{eq:sampleconprob}
    & \prod_{\{i,j\}\in\mathcal{E}_H}\min\Big(\frac{n^{\alpha_i}n^{\alpha_j}}{\mu n n^{d\beta_{\{i,j\}}}},1\Big)^{1/T} \nonumber\\
    & = n^{\sum_{\{i,j\}\in\mathcal{E}_H}1/T\min(\alpha_i + \alpha_j - d \beta_{\{i,j\}} - 1, 0)}.
\end{align}
Combining~\eqref{eq:sampleweights},~\eqref{eq:samplepositions} and~\eqref{eq:sampleconprob} yields that the expected number of subgraphs $H$ with  weights proportional to $n^{\alpha_i}$, $i\in[k]$ and distances proportional to $n^{\beta_{\{i,j\}}}$ $i,j\in[k]$ scales as
\begin{align}
    & n^{k + (1 - \tau) \sum_{i=1}^{k} \alpha_i + d \sum_{j=2}^{k} \min_{i:i<j} \beta_{\{i,j\}}}\nonumber\\
    & \times n^{ 1/T \sum_{ (i,j) \in \mathcal{E}} \min(\alpha_i + \alpha_j - d \beta_{\{i,j\}} - 1, 0)}.
\end{align}
If this exponent is maximized at a unique set of $\boldsymbol\alpha$ and $\boldsymbol\beta$ vectors, then their contribution to the expected number of subgraphs $H$ scales polynomially larger than the contribution of any other subgraph configuration, implying that this configuration dominates the subgraph count of $H$. We therefore optimize the exponent over all allowed $\boldsymbol{\alpha}$ and $\boldsymbol{\beta}$, resulting in the optimization problem
  \begin{align} \label{eq:optproblem}
     f^l(H):=&\max_{\substack{\boldsymbol{
    \alpha}\in[0,\frac{1}{\tau-1}]^k,\\\boldsymbol{
    \beta}\in[-\frac{1}{d},0]^k}}  k + (1 - \tau) \sum_{i=1}^{k} \alpha_i + d \sum_{j=2}^{k} \min_{i:i<j} \beta_{\{i,j\}} \nonumber\\
    & + 1/T \sum_{ (i,j) \in \mathcal{E}} \min(\alpha_i + \alpha_j - d \beta_{\{i,j\}} - 1, 0)\nonumber\\
     \text{s.t. } \quad &
         \beta_{\{i,j\}} \leq \max(\beta_{\{i,s\}},\beta_{\{s,j\}})
        \quad \forall i\neq j \neq s \in [k].  
  \end{align}
  Although the optimization problem is not linear in $\boldsymbol\alpha$ and $\boldsymbol\beta$ due to the max and min functions, it can be rewritten as a mixed integer linear program (MILP) (see Appendix~\ref{ap:milp}), which can be solved efficiently in Gurobi~\cite{gurobi}. Then, if~\eqref{eq:optproblem} has a unique solution, 
  \begin{equation}\label{eq:NHgirg}
  	N^{GIRG}(H)\sim n^{f^l(H)},
  \end{equation}
  so that it characterizes the scaling of subgraph $H$ in the geometric random graph model. 

  While~\eqref{eq:optproblem} has the parameter $1/T$ in its objective, the optimal $\boldsymbol\alpha$ and $\boldsymbol\beta$ variables
 are independent of the inverse temperature $
  1/T$~\cite{michielan2024optimal}. Indeed, all optimal subgraph configurations minimize `edge-energy', and satisfy
  \begin{equation}\label{eq:edgenergy}
      \alpha_i + \alpha_j - d \beta_{\{i,j\}} = 1 \quad \forall \{i,j\}\in\mathcal{E}_H.
  \end{equation}
  This constraint means that in any optimal subgraph configuration, the connection probability~\eqref{eq:conproblow} scales as a constant in $n$. In other words, the subgraph is most likely formed on vertices that are of sufficiently high weight, or sufficiently close such that they are likely to connect. 
Using~\eqref{eq:edgenergy} enables us to rewrite~\eqref{eq:optproblem} as
  \begin{align} \label{eq:optproblemreduced}
     f^l(H):=&\max_{\substack{\boldsymbol{
    \alpha}\in[0,\frac{1}{\tau-1}]^k,\\\boldsymbol{
    \beta}\in[-\frac{1}{d},0]^k}}  k + (1 - \tau) \sum_{i=1}^{k} \alpha_i + d \sum_{j=2}^{k} \min_{i:i<j} \beta_{\{i,j\}} \nonumber\\
     \text{s.t. } \quad &
         \beta_{\{i,j\}} \leq \max(\beta_{\{i,s\}},\beta_{\{s,j\}})
        \quad \forall i\neq j \neq s \in [k].   \nonumber\\ 
        &  \alpha_i + \alpha_j - d \beta_{\{i,j\}} =1 \quad \forall (i,j) \in \mathcal{E_H},
  \end{align}
  which is indeed $1/T$-independent. Then~\eqref{eq:NHgirg} implies that for all subgraphs $H$, the scaling of the subgraph count in $n$ is independent of the inverse temperature $1/T$ as long as $1/T>1$. 
  
\section{Motif optimization: $1/T<1$ }\label{sec:motifhightemp}
We now turn to the weak geometric or high-temperature regime, where $1/T<1$, and construct a similar optimization problem to~\eqref{eq:optproblem}. The probability that a given set of $k$ randomly selected vertices has weights proportional to $(n^{\alpha_i})_{i\in[k]}$ and distances proportional to $(n^{\beta_{ij}})_{ij\in V_H\times V_H}$ does not depend on the connection probability, and therefore does not depend on $1/T$. Therefore, ~\eqref{eq:sampleweights} and~\eqref{eq:samplepositions} still hold, and the only thing that is left is to compute the probability that vertices with weights proportional to $n^{\alpha_i}$ $i\in[k]$ and distances proportional to $n^{\beta_{\{i,j\}}}$ $i,j\in[k]$ form subgraph $H$, similarly to~\eqref{eq:sampleconprob}. 
Using~\eqref{eq:conprobhigh}, we obtain that when $1/T<1$,  the probability that vertices with weights proportional to $n^{\alpha_i}$ $i\in[k]$ and distances proportional to $n^{\beta_{\{i,j\}}}$ $i,j\in[k]$ form subgraph $H$ scales as
\begin{align}\label{eq:sampleconprobhightemp}
    & \prod_{\{i,j\}\in\mathcal{E}_H}\min\Big(\frac{n^{\alpha_i}n^{\alpha_j}}{\mu n n^{ d\beta_{\{i,j\}}/T},1}\Big) \nonumber\\
    & = n^{\sum_{\{i,j\}\in\mathcal{E}_H}\min(\alpha_i + \alpha_j - d \beta_{\{i,j\}}/T - 1, 0)}.
\end{align}
Combining this with~\eqref{eq:sampleweights} and~\eqref{eq:samplepositions} in a similar fashion as~\eqref{eq:optproblem} leads to the optimization problem
 \begin{align} \label{eq:optproblemhightemp}
    f^h(H):=  & \max_{\substack{\boldsymbol{
    \alpha}\in[0,\frac{1}{\tau-1}]^k,\\\boldsymbol{
    \beta}\in[-\frac{1}{d},0]^k}}  k + (1 - \tau) \sum_{i=1}^{k} \alpha_i + d \sum_{j=2}^{k} \min_{i:i<j} \beta_{\{i,j\}} \nonumber\\
    & + \sum_{ (i,j) \in \mathcal{E}} \min(\alpha_i + \alpha_j - d \beta_{\{i,j\}}/T - 1, 0)\\
     \text{s.t. } \quad &
         \beta_{\{i,j\}} \leq \max(\beta_{\{i,s\}},\beta_{\{s,j\}})
        \quad \forall i\neq j \neq s \in [k].  \label{eq:optproblem_normal_constraint}
  \end{align}
When $1/T\to 0 $, this optimization problem is equivalent to the one for Chung-Lu random graphs~\cite{stegehuis2019}, which essentially sets all $\beta_{\{ij\}}$ to zero, as the Chung-Lu model does not contain positional variables, only the $\alpha$ variables corresponding to the vertex weights. We will denote the solution of~\eqref{eq:optproblemhightemp} with $1/T=0$ by $f^h_{1/T=0}(H)$, which provides the scaling of the subgraph count $H$ in the Chung-Lu model~\ref{eq:conprobcl}.

Therefore, when the optimizer of~\eqref{eq:optproblemhightemp} is attained for $\beta_{ij}=0$ for all $i,j$, the optimizer is the same as the one for the Chung-Lu version $f^h_{1/T=0}(H)$. In that case, the particular subgraph is most dominantly formed by vertices of specific degrees at any inter-distance. It also implies that the subgraph occurs in the same order of magnitude in a geometric model as in its non-geometric counterpart, and is ineffective in distinguishing geometric random graph models from non-geometric ones. 
On the other hand, when $\beta_{ij}>0$ in the unique optimal solution for some $i,j$, then $f^h(H)>f^h_{1/T=0}(H)$, which means that the subgraph count of $H$ scales at a larger exponent in $n$ in the geometric model~\ref{eq:conprobhigh} than in the non-geometric counterpart~\eqref{eq:conprobcl}, and the subgraph can effectively distinguish the two models. 

When $1/T\to 0$, the optimizer of~\eqref{eq:optproblemhightemp} tends to $\beta_{ij}=0$ for all subgraphs $H$. 
That means that for all subgraphs, there is some transition point $1/T_c\geq 0$ where the optimal subgraph structure of~\eqref{eq:optproblemhightemp} changes from the `geometric regime' of $f^h(H)>f^h_{1/T=0}(H)$ to being non-geometric with all $\beta_{ij}=0$ and $f^h(H)=f^h_{1/T=0}(H)$.  

  \section{Optimal subgraph to detect geometry}\label{sec:optmotif}
  We now investigate which subgraph is the most powerful indicator of the presence of geometry. In particular, we investigate which statistic distinguishes the Chung-Lu model and the GIRG the most. That is, we maximize
\begin{align}\label{eq:BH}
   B(H) &:= \log\Big(\frac{N^{GIRG}(H)}{N^{IRG}(H)}\Big)/\log(n)\nonumber\\
   &  = f(H)-f^h_{1/T=0}(H),
\end{align}
among all subgraphs $H$ of a specific size. 
Here
\begin{equation}
	f(H) = \begin{cases}
		f^l(H) & 1/T>1\\
		f^h(H) & 1/T<1.
	\end{cases}
\end{equation}
Then $B(H)$ indicates the difference in scaling of the occurrence of subgraph $H$ in the GIRG and the Ching-Lu model. Thus, when $B(H)>0$, a subgraph occurs at a larger scale in the network size $n$ in GIRG compared to the Chung-Lu model, and the subgraph effectively distinguishes the GIRG and the Chung-Lu model. 

\begin{figure}
    \centering
    \includegraphics[width=\linewidth]{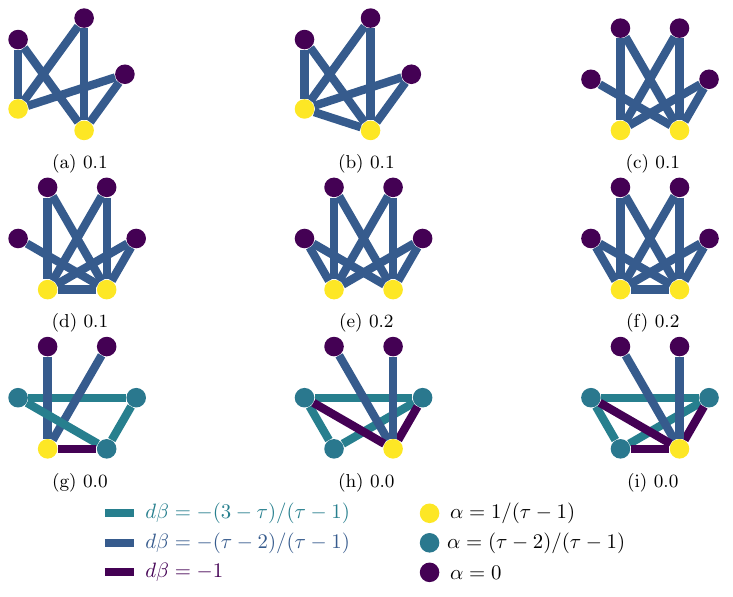}
    \caption{$B(H)$ for all subgraphs of size 4, 5, 6 with $B(H)>0$ for $\tau=2.4, 1/T>1$. The colors of the nodes and edges denote the optimal $\boldsymbol{\alpha}$ and $\boldsymbol{\beta}$ values of~\eqref{eq:optproblem}.}
    \label{fig:tau24lowunweighted}
\end{figure}

Figure~\ref{fig:tau24lowunweighted} presents all subgraphs with $B(H)>0$ for subgraphs on 4, 5, and 6 nodes for $\tau=2.4$ and $1/T>1$. Interestingly, the 'mountains subgraph' (Figure~\ref{fig:tau24lowunweighted}(f)), triangles that are glued together at a single edge, is the most powerful subgraph. While cliques are generally used as evidence for geometry, this subgraph differentiates the two models better. The figure also shows that in the geometric model, these subgraphs are typically formed on two hubs, and three low-degree vertices. Furthermore, all vertices then typically appear at distance proportional to $n^{-(\tau-2)/(\tau-1)}$, which is closer than the maximal possible distance of $n^{0}$. Other subgraphs, such as Figure~\ref{fig:tau24lowunweighted}(g) contain pairs of vertices that appear at different distances. 

\begin{figure}
    \centering
    \includegraphics[width=\linewidth]{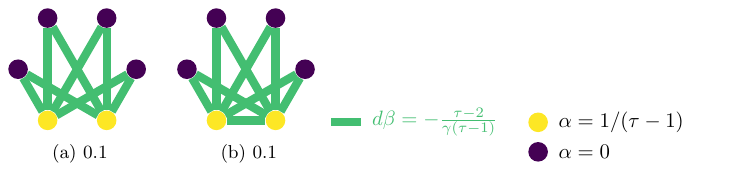}
    \caption{$B(H)$ for all subgraphs of size 4, 5, 6 with $B(H)>0$ for $\tau=2.4, 1/T=0.95$. The colors of the nodes and edges denote the optimal $\boldsymbol{\alpha}$ and $\boldsymbol{\beta}$ values of~\eqref{eq:optproblemhightemp}. }
    \label{fig:tau24highunweightedgamma09}
\end{figure}

Figure~\ref{fig:tau24highunweightedgamma09} plots all subgraphs of size 4, 5, 6 with $B(H)>0$ for $1/T<1$. In this case, already for $1/T=0.95$, the number of subgraphs that can detect geometry has decreased significantly compared to Figure~\ref{fig:tau24lowunweighted}. Again, interestingly, the mountains subgraph (Figure~\ref{fig:tau24highunweightedgamma09}(b)) is one of the most powerful subgraphs to detect geometry, but compared to Figure~\ref{fig:tau24lowunweighted}, the value of $B(H)$ has dropped from 0.2 to 0.1, meaning that in the high-temperature regime, the difference between the number of times this subgraph appears in a geometric model and a Chung-Lu counterpart is smaller than in the low-temperature regime. Furthermore, compared to Figure~\ref{fig:tau24lowunweighted}, the typical distance between the vertices in the mountains subgraph has increased from $n^{-(\tau-2)/(\tau-1)}$ to $n^{-T(\tau-2)/(\tau-1)}$. Intuitively, nearby connections give stronger evidence for geometry. However, nearby connections are more rare than further connections. Due to the additional randomness of the increase in temperature, these nearby connections become more rare and the subgraphs are more likely to form between higher-degree vertices. 


\paragraph{Most powerful subgraph: mountains subgraph.}
We now investigate the fundamental limit of using subgraph statistics to detect network geometry. To that end, we investigate the `mountains' subgraph on $k$ vertices, $M_k$, as the optimal subgraph statistic in more detail. In the non-geometric Chung-Lu model, the optimizer of~\eqref{eq:optproblem} for $1/T=0$ is given by
\begin{equation}\label{eq:mountainchunglu}
    f^h_{1/T=0}(M_k)=k-2+(2-\tau)(k-2).
\end{equation}
This optimizer is attained by setting $\alpha_i=(\tau-2)/(\tau-1)$ for $i=1,\dots, k-2$, and $\alpha_i=1/(\tau-1)$ for $i=k-1,k$. Therefore, this subgraph typically appears on $k-2$ vertices of degree proportional to $n^{(\tau-2)/(\tau-1)}$ and two vertices of degree proportional to $n^{1/(\tau-1)}$. 
In the geometric regime for $1/T>1$ on the other hand, for large $\tau$, the optimizer of~\eqref{eq:optproblem} equals
\begin{equation}\label{eq:mountain}
    f^l(M_k) = k-2+(k-1)\frac{2-\tau}{\tau-1}.
\end{equation}
In this case, the optimizer is attained by $k-2$ vertices with $\alpha_i=0$ and two with $\alpha_i=1/(\tau-1)$, while $\beta d = (2-\tau)/(\tau-1)$ for all pairs of nodes. We now compute the transition point in $\tau$ at which~\eqref{eq:mountainchunglu} becomes larger than~\eqref{eq:mountain}. That is, we compute the limit of $\tau$ underneath which $M_k$ counts scale the same in geometric models as in non-geometric ones, which is as long as $f^l(M_k)- f^h_{1/T=0}(M_k)>0$, which translates to
\begin{equation}
    \tau > 1+\frac{k-1}{k-2}.
\end{equation}
Thus, when $1/T>$, then for every $\tau>2$ there exists a value of $k$ such that the $k$-mountains subgraph can distinguish geometric and non-geometric random graphs. This in contrast to cliques, which turn out to be incapable of distinguishing geometric and non-geometric networks in this regimes for $\tau<7/3$~\cite{michielan2022,blasius2018b}.

We now turn to the weak goemetry regime $1/T<1$. In that case, the optimizer of~\eqref{eq:optproblemhightemp} for $1/T$ close to one equals
\begin{equation}\label{eq:mountaingeom}
    f^h(M_k) = -2+(k-1)\frac{2-\tau}{(\tau-1)/T}.
\end{equation}
The optimizer in this case is attained by $k-2$ vertices with $\alpha_i=0$ and two with $\alpha_i=1/(\tau-1)$, while $\beta d = (2-\tau)/((\tau-1)/T)\in(-1,0)$. We now again compute the transition point in $1/T$ where~\eqref{eq:optproblemhightemp} is optimized by the same optimizer as the Chung-Lu optimizer~\eqref{eq:mountainchunglu}. This yields that the phase transition in $T$ where $f^h(M_k)-f^h_{1/T=0}(M_k)=0$ occurs at
\begin{equation}
    1/T_c = \frac{k-1}{(k-2) (\tau -1)}.
\end{equation}
When $1/T<1/T_c$, then the $M_k$ count scales similarly in the Chung-Lu model and the geometric inhomogeneous random graph, which means that even the best performing subgraph is not capable of detecting geometry in this case. The phase transition $1/T_c$ decreases when $k$ increases. When $k\to\infty$, the phase transition tends to $1/(\tau-1)$. For example, this means that the fundamental limit of subgraph counts to detect geometry is at $1/(\tau-1)$.

\paragraph{Cliques.}
We now show that cliques are less efficient in detecting geometry. When $1/T>1$, it is known that $k$-cliques are an effective distinguisher as long as $k <2/(3-\tau)$~\cite{michielan2021}. 

When $1/T<1$, there are two competing potential optimizers of~\eqref{eq:optproblemhightemp}. The first is the non-geometric optimizer which also optimizes the Chung-Lu model, $\beta_{ij}=0$ for all $i,j$ and $\alpha_i=1/2$ for all $i$, resulting in
\begin{equation}
    f^h_{1/T=0}(K_k) = k+(1-\tau)k/2.
\end{equation}
The second is the the geometric optimizer, $\beta_{ij}=-1/d$ for all $i,j$ and $\alpha_i=(1-1/T)/2$ for all $i$, which gives for $1/T$ close to one
\begin{equation}
    f^h(K_k) = k-(k-1)+k(\tau-1)(1/T-1)/2.
\end{equation}


Therefore, gives that the transition value $1/T_c$ where $f^h(K_k)-f^h_{1/T=0}(K_k)= 0$ equals
\begin{equation}
    1/T_c = \frac{2 (k-1)}{k (\tau -1)}.
\end{equation}
This transition value is above one when $\tau<3-2/k$. Thus, in that regime, cliques do not distinguish geometric and non-geometric graphs for any $1/T<1$. Furthermore, $1/T_c$ is decreasing in $k$, so that the lowest $1/T$ value at which geometry is detectable is from triangles, at $1/T_c=4/(3(\tau-1))$, putting the fundamental limit of triangle counts to detect geometry at $1/T_c=4/(3(\tau-1))$, which is higher than the limit $1/(\tau-1)$ from the mountains subgraph $M_k$.

\section{Most powerful degree-based subgraph}\label{sec:optweightedmotif}
We now show that while subgraph counts can only detect geometry up to $1/T=1/(\tau-1)$, different types of statistics are more powerful to detect geometry. In particular, we will turn to statistics of the type
\begin{equation}\label{eq:Nha}
    N_H(\boldsymbol{\alpha}) = \sum_{v_1,\dots,v_k \in [n]^k}\mathbbm{1}_{v_1,\dots,v_k \text{ form } H}\prod_{i\in[k]}\mathbbm{1}_{d_{v_i}\leq n^{\alpha_i}}.
\end{equation}
That is, we only count subgraphs of type $H$ of which the degrees are at most $n^{\alpha_i}$ for $i\in[k]$.
We are interested in obtaining the statistic that distinguishes the Chung-Lu model and the GIRG the most, and maximize
\begin{equation}\label{eq:BWa}
    B_W(H):= \max_{\boldsymbol{\alpha}}\log\Big(\frac{N_H^{GIRG}(\boldsymbol{\alpha})}{N_H^{IRG}(\boldsymbol{\alpha})}\Big)/\log(n),
\end{equation}
similarly to~\eqref{eq:BH}. 
However, for a statistic in the form of~\eqref{eq:Nha} to be useful on a network data set, the particular subgraph should also appear sufficiently often in the network. That is, we require that $N_H^{GIRG}(\boldsymbol{\alpha})\gg 1$. 

Optimizing~\eqref{eq:BWa} corresponds to optimizing the difference between the objective of~\eqref{eq:optproblemhightemp} for a specific value of $T$ and the objective with $1/T = 0$. In contrast to~\eqref{eq:BH}, we subtract the objectives, and then optimize over $\alpha$ and $H$ afterwards. This results in the optimization problem

 \begin{align} \label{eq:optproblemweighted}
     \max_{\substack{\boldsymbol{
    \alpha}\in[0,\frac{1}{\tau-1}]^k,\\\boldsymbol{
    \beta}\in[-\frac{1}{d},0]^k}} & d \sum_{j=2}^{k} \min_{i:i<j} \beta_{\{i,j\}}  - \sum_{ (i,j) \in \mathcal{E}} \min(\alpha_i + \alpha_j - 1, 0)\nonumber\\
    & + \sum_{ (i,j) \in \mathcal{E}_H} \min(\alpha_i + \alpha_j -  d \beta_{\{i,j\}}/T  - 1, 0)\\
     \text{s.t. } \quad &
     k + (1 - \tau) \sum_{i=1}^{k} \alpha_i + d \sum_{j=2}^{k} \min_{i:i<j} \beta_{\{i,j\}} \nonumber\\
    & + \sum_{ (i,j) \in \mathcal{E}_H} \min(\alpha_i + \alpha_j -  d \beta_{\{i,j\}}/T - 1, 0)\geq 0 \nonumber\\
         &\beta_{\{i,j\}} \leq \max(\beta_{\{i,s\}},\beta_{\{s,j\}})
        \quad \forall i\neq j \neq s \in [k].  \label{eq:optproblem_normal_constraint}
  \end{align}
A similar optimization problem in which $\min(\alpha_i + \alpha_j -  d \beta_{\{i,j\}}/T  - 1, 0)$ is replaced by $\min(\alpha_i + \alpha_j - d \beta_{\{i,j\}}  - 1, 0)/T$ yields the optimal statistic for $1/T>1$. Again, we show in Appendix~\eqref{ap:milp} that this optimization problem can be converted to a mixed-integer linear program. 

\begin{figure}
    \centering
    \includegraphics[width=\linewidth]{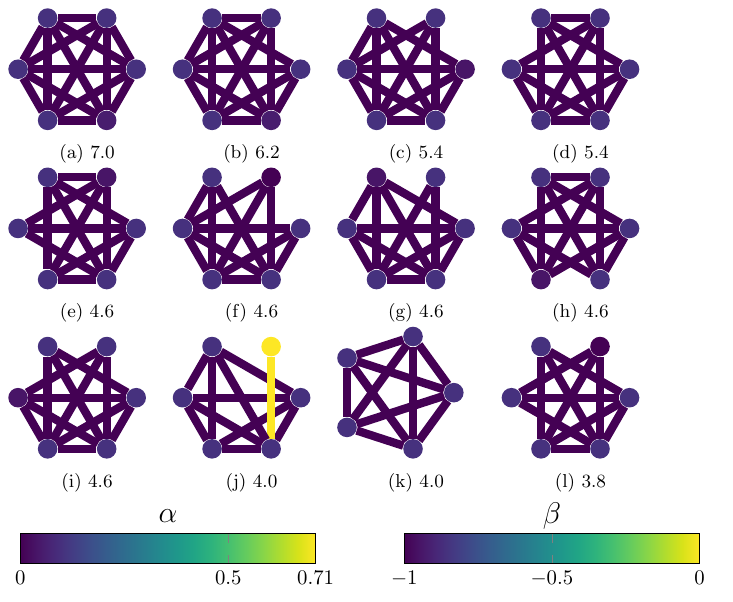}
    \caption{$B_W(H)$ for all subgraphs of size 4, 5, 6 with $B_W(H)>0$ for $\tau=2.4, 1/T=0.8$. The colors of the nodes and edges denote the optimal $\boldsymbol{\alpha}$ and $\boldsymbol{\beta}$ values of~\eqref{eq:optproblemweighted}.}
    \label{fig:tau24highweightedgamma08}
\end{figure}
\begin{figure}
    \centering
    \includegraphics[width=\linewidth]{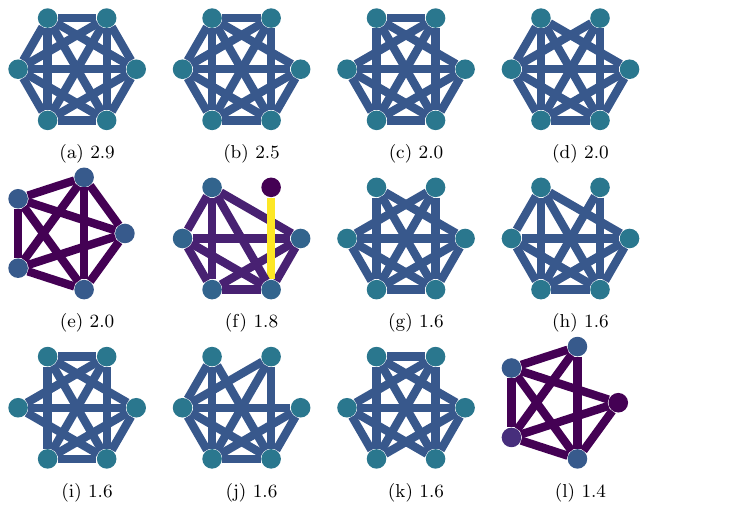}
    \caption{$B_W(H)$ for the 12 subgraphs of size 4, 5, 6 with the largest values of $B_W(H)>0$ for $\tau=2.2, 1/T=0.6$. The colors of the nodes and edges denote the optimal $\boldsymbol{\alpha}$ and $\boldsymbol{\beta}$ values of~\eqref{eq:optproblemweighted}, for which the legend is as in Figure~\ref{fig:tau24highweightedgamma08}.}
    \label{fig:tau24highweightedgamma06}
\end{figure}
\begin{figure}
    \centering
    \includegraphics[width=\linewidth]{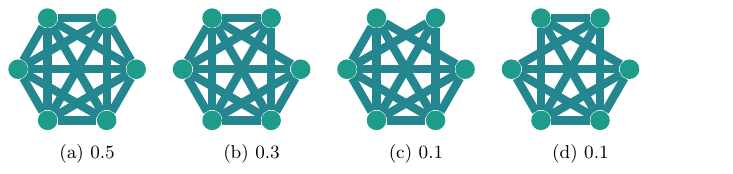}
    \caption{$B_W(H)$ for the 12 subgraphs of size 4, 5, 6 with the largest values of $B_W(H)>0$ for $\tau=2.2, 1/T=0.4$.The colors of the nodes and edges denote the optimal $\boldsymbol{\alpha}$ and $\boldsymbol{\beta}$ values of~\eqref{eq:optproblemweighted}, for which the legend is as in Figure~\ref{fig:tau24highweightedgamma08}.}
    \label{fig:tau24highweightedgamma04}
\end{figure}

Figures~\ref{fig:tau24highweightedgamma08},~\ref{fig:tau24highweightedgamma06} and~\ref{fig:tau24highweightedgamma04} show the subgraphs of sizes 4, 5 and 6 for which $B_W(H)>0$. These figures show that the smaller $\gamma$, the fewer degree-based subgraphs are able to detect geometry. Furthermore, for lower $\gamma$, the degrees in the degree-based subgraphs increase, as well as the typical distances between these vertices. Furthermore, while regular subgraphs only had a low distinguishing value $B(H)$ for quite high $\gamma$ (see Figure~\ref{fig:tau24highunweightedgamma09}), Figure~\ref{fig:tau24highweightedgamma04} shows that using these degree-based subgraphs is much more effective: they are able to detect geometry up to much higher temperatures and have a higher distinguishing power $B_W(H)$. 


\paragraph{Optimal weighted statistic: weighted cliques}
Figures~\ref{fig:tau24highweightedgamma08},~\ref{fig:tau24highweightedgamma06} and~\ref{fig:tau24highweightedgamma04} show that the clique is the subgraph with the highest distinguishing power $B_W(H)$. For the $k$-clique, the optimal solution to~\eqref{eq:optproblemweighted} is the $k$-clique where all vertices have an equal $\alpha$, equaling
\begin{equation}\label{eq:optalph}
    \alpha = \frac{k(1-1/T)-1}{k((1-\tau)/T+2)-2},
\end{equation}
as long as $1/T>2/k$. Below $2/k$, no subgraph-based statistic of size $k$ can effectively distinguish geometry. Also, note that this $\alpha$ is always smaller or equal to $1/(\tau-1)$ as long as $\tau<3$.

Indeed, all nodes are isomorphic in a clique, so that the optimal solution has equal $\alpha$ and $\beta$ values for all nodes. Then, ~\eqref{eq:optproblemweighted} becomes
\begin{align} \label{eq:optproblemrenormalized}
    & \max_{\alpha,\beta}\frac{k(k-1)}{2}[\min(2 \alpha - 1 - d \beta /T, 0) + \max(0, 1 - 2 \alpha)]  \nonumber\\
    & + (k-1)d\beta \nonumber\\
    & \text{s.t. }k + k(1-\tau)\alpha + (k-1)d\beta \nonumber\\
    & + \min(2 \alpha - 1 - d \beta /T, 0)k(k-1)/2 \geq 0
\end{align}
where $\alpha \in [0,1/(\tau-1)]$, $\beta \in [-1/d,0]$. As this is a piece-wise linear optimization problem, the optimizer is either at one of the boundary points for $\alpha$ and $\beta$, or at one of the `kinks', $ 2\alpha-1=d\beta/T$, or $\alpha=1/2$. Note that the boundary points for $\alpha$ and $\beta$ include the boundary prescribed by the constraint of the optimization problem. Computing these optimizers shows that the optimal solution is given by the $\alpha$ and $\beta$ values prescribed by the constraint. For low $\alpha$, the $2\alpha$ and $-2\alpha$ in the minima in the objective cancel each other out. The $d\beta$ term occurs with a negative coefficient, which leads to the fact that $\beta$ should be as small as possible. However, this only holds when $\alpha$ can still be sufficiently small so that $2 \alpha - 1 - d \beta /T<0$. Thus, we have to set $2\alpha-1 = d\beta/T$. Plugging this into the constraint results in~\eqref{eq:optalph}, as long as this solution still satisfies the boundary requirements $\alpha\geq 0 $ and $\beta\geq -1/d$, which is as long as
\begin{equation}
1/T< \frac{k \tau -k-2}{k (\tau -1)}.
\end{equation}
In this setting, the objective function becomes
\begin{equation}
    B_W(K_k)= (k-1)d\beta \big(1-\frac{k}{2T}\big).
\end{equation}
As the optimizer has $\beta<0$, this indicates that $B_W(K_k)>0$ as long as $1/T>2/k$. Interestingly, this means that for any $T<\infty$, it is possible to choose a $k$ such that these degree-weighted $k$-cliques are able to distinguish weak geometry from no geometry, in contrast to using pure subgraph counts, which have a fundamental threshold at $1/T=1/(\tau-1)$. 

When $1/T>\frac{k \tau -k-2}{k (\tau -1)}$ the optimal solution is $\beta=-1/d$ and any $\alpha\in[\frac{k (1/T -  k/T+k-1)-2}{2 k (k-\tau )},(1/T-1)/2]$ is optimal. 

For $1/T>1$, the corresponding problem to~\eqref{eq:optproblemrenormalized} for cliques is
\begin{align} \label{eq:optproblemrenormalizedlow}
    & \max_{\alpha,\beta}\frac{k(k-1)}{2}[\min(2 \alpha - 1 - d \beta , 0)/T + \max(0, 1 - 2 \alpha)]  \nonumber\\
    & + (k-1)d\beta \nonumber\\
    & \text{s.t. }k + k(1-\tau)\alpha + (k-1)d\beta \nonumber\\
    & + \min(2 \alpha - 1 - d \beta , 0)k(k-1)/(2T) \geq 0.
\end{align}
Here, the optimizer tries to achieve $\alpha$ and $\beta$ as low as possible, and the solution at the boundary $\alpha=0,\beta=-1/d$ is feasible, making it the optimal solution. This also demonstrates the optimality of earlier described statistics designed specifically for the low-temperature regime~\cite{michielan2022,friedrich2024}.

\section{Conclusion}\label{sec:conclusion}
In this paper, we have explored the most effective subgraph and degree-based statistics for distinguishing between geometric and non-geometric power-law random graphs, particularly in the weak geometry regime. Our findings demonstrate that, regardless of the level of weak geometry, subgraph and degree-based statistics can effectively differentiate between geometric and non-geometric random graphs. Interestingly, while cliques alone are insufficient for this distinction, counting specific cliques of certain maximal degrees proves to be effective.

Future research could investigate the performance of our method on other types of kernels, such as those generated by spatial preferential attachment or similar models~\cite{jacob2015}.

Furthermore, we believe that the Mixed-Integer Linear Programming (MILP) approach that we introduce here has the potential to capture a broader range of network statistics beyond~\eqref{eq:Nha}. Specifically, the MILP can easily be reformulated to analyze statistics that are based on local information, such as clustering coefficients, and we believe that it is also amenable to analyzing subgraph-based embeddings such as node2vec, or even spectral methods. 

It would be interesting to see whether such other types of statistics, such as those based on eigenvalues or centered subgraph counts could achieve similar results. Our current work is limited to subgraphs, but applying a similar approach to induced subgraphs could yield better-performing statistics.

Whereas the results for $1/T>1$ are $1/T$-independent, for $1/T<1$, the optimal statistic for distinguishing geometric and non-geometric random graphs it does depend on $1/T$. This suggests that the inverse temperature can be estimated using degree-based subgraph statistics, presenting an intriguing direction for future research.

\paragraph*{Acknowledgements.}
CS is supported by an NWO VENI grant 202.001.

\bibliography{references}

\appendix
\section{Mixed Integer Linear Programming formulations}\label{ap:milp}
The optimization program~\eqref{eq:optproblem} can be rewritten as a Mixed integer linear program (MILP) in the following way
\begin{subequations}
\begin{alignat}{3}
    &\max  \sum_{u\in V}(1-\tau)\alpha_v+d\zeta_u+\sum_{u,v\in\mathcal{E}_H}&&\delta_{u,v}/T\\
    &\text{s.t. } \delta_{u,v}+d\beta_{u,v}\leq \alpha_u+\alpha_v &&\forall \{u,v\}\in\mathcal{E}_H\label{eq:delta1}\\
    & \zeta_u\leq \beta_{u,v} && \forall u<v\\
    & \delta_{u,v}\leq 0 && \forall \{u,v\}\in\mathcal{E}_H \label{eq:delta2}\\
    & \zeta_u \geq -1/d &&\forall u\\
    & \kappa_{u,v,w}+\kappa_{v,u,w} = 1 && \forall u,v,w\label{eq:kappasum}\\
    & \beta_{u,v}\leq \beta_{u,w}+\kappa_{u,v,w}/d  &&\forall u,v,w\label{eq:trianglecnstr}\\
    & \kappa_{u,v,w}\in\{0,1\} && \forall u,v,w.
\end{alignat}
\end{subequations}
Here $\delta_{uv}$ encodes $\min(\alpha_u+\alpha_v-d\beta_{u,v}-1,0)$ with the constraints~\eqref{eq:delta2} and~\eqref{eq:delta2}, as this term appears in the maximizer objective. The term $\zeta_u$ encodes the $\min_{v:u<v}\beta_{u,v}$ term. Finally, the $\kappa_{u,v,w}$ are auxiliary binary variables that enable to encode the triangle inequality constraint $\beta_{u,v}\leq \max(\beta_{u,w},\beta_{v,w})$, and encodes whether $\beta_{vw}\leq \beta_{uw}$. Indeed, as $\beta_{u,v}\in [-1/d,0]$, when $\kappa_{u,v,w} = 1$, the constraint~\eqref{eq:trianglecnstr} always holds. In that case~\eqref{eq:kappasum} ensures that $\kappa_{v,u,w}= 0$, and thus $\beta_{u,v}\leq \beta_{v,w}$.

Very similarly, the MILP corresponding to~\eqref{eq:optproblemhightemp} becomes
\begin{subequations}
\begin{alignat}{3}
    &\max  \sum_{u\in V}(1-\tau)\alpha_v+d\zeta_u+\sum_{u,v\in\mathcal{E}_H}&&\delta_{u,v}/T\\
    &\text{s.t. } \delta_{u,v}+d\beta_{u,v}/T\leq \alpha_u+\alpha_v &&\forall \{u,v\}\in\mathcal{E}_H\\
    & \zeta_u\leq \beta_{u,v} && \forall u<v\\
    & \delta_{u,v}\leq 0 && \forall \{u,v\}\in\mathcal{E}_H \\
    & \zeta_u \geq -1/d &&\forall u\\
    & \kappa_{u,v,w}+\kappa_{v,u,w} = 1 && \forall u,v,w\\
    & \beta_{u,v}\leq \beta_{u,w}+\kappa_{u,v,w}/d  &&\forall u,v,w\\
    & \kappa_{u,v,w}\in\{0,1\} && \forall u,v,w.
\end{alignat}
\end{subequations}

The MILP corresponding to~\eqref{eq:optproblemweighted} (for $1/T<1$) can be written as
\begin{subequations}
\begin{alignat}{3}
    &\max  d\zeta_u+\sum_{u,v\in\mathcal{E}_H}(\delta_{u,v}-\hat{\delta}_{u,v})/T&&\\
    &\text{s.t. } \delta_{u,v}+d\beta_{u,v}\leq \alpha_u+\alpha_v &&\forall \{u,v\}\in\mathcal{E}_H\\
    & \zeta_u\leq \beta_{u,v} && \forall u<v\\
    & \delta_{u,v}\leq 0 && \forall \{u,v\}\in\mathcal{E}_H \\
    & \zeta_u \geq -1/d &&\forall u\\
    & \kappa_{u,v,w}+\kappa_{v,u,w} = 1 && \forall u,v,w\\
    & \beta_{u,v}\leq \beta_{u,w}+\kappa_{u,v,w}/d  &&\forall u,v,w\\
    & \hat{\delta}_{u,v}\geq \alpha_u+\alpha_v-1-\theta_{u,v} && \forall u,v\in\mathcal{E}_H\label{eq:theta1}\\
    & \hat{\delta}_{u,v}\geq \theta_{u,v}-1&&\forall u,v\in\mathcal{E}_H\label{eq:theta2}\\
    &  k +  \sum_{i=1}^{k} (1 - \tau)\alpha_i + d \zeta_i + \sum_{ (i,j) \in \mathcal{E}_H} \delta_{i,j} && \geq  0 \\
    & \kappa_{u,v,w}\in\{0,1\} && \forall u,v,w\\
    & \theta_{u,v}\in\{0,1\}&& \forall u,v \in\mathcal{E}_H.
\end{alignat}
\end{subequations}
Here $\hat{\delta}_{u,v}$ encodes $\min(\alpha_u+\alpha_v-1,0)$. As this appears with a negative sign in the objective function (which has a maximization objective), we cannot encode this non-linear function in the same way as $\delta_{u,v}$, which appears with a positive factor. Indeed, as the objective aims to maximize $\delta_{u,v}$, constraining it to be smaller than the two terms in the minimum suffices to make it achieve the minimum. For $\hat{\delta}_{u,v}$, which is minimized, this would not work, as the MILP would then make $\hat{\delta}_{uv}$ as small as possible. Therefore, we introduce the binary variables $\theta_{uv}$ which encodes whether $\alpha_u+\alpha_v-1<0$ or not. When $\theta_{u,v}=1$, then~\eqref{eq:theta2} becomes $\hat{\delta}_{u,v}\geq 0$, while~\eqref{eq:theta1} becomes $\hat{\delta}_{u,v}\geq \alpha_u+\alpha_v-2$. As $\alpha_u,\alpha_v\in[0,1]$, the latter is always satisfied when $\hat{\delta}_{u,v}\geq 0$. Furthermore, because $\hat{\delta}_{u,v}$ appears negatively in the objective function, it will set  $\hat{\delta}_{u,v}=0$. Thus, $\theta_{u,v}=1$ encodes that $\alpha_u+\alpha_v-1>0$ and sets $\hat{\delta}_{u,v}\geq 0$. 

On the other hand, when $\theta_{u,v}= 0$, then constrain~\eqref{eq:theta1} becomes $\hat{\delta}_{u,v}\geq \alpha_u+\alpha_v-1$ and~\eqref{eq:theta2} becomes $\hat{\delta}_{u,v}\geq -1$. In this case, the first constraint dominates, and due to the negative coefficient of $\hat{\delta}_{u,v}$, the maximizer will set $\hat{\delta}_{u,v}= \alpha_u+\alpha_v-1$.

\end{document}